\documentclass{appolb}
\usepackage{epsfig}

\begin{document}
\title{Resonance Production in Au+Au and p+p Collisions at $\sqrt{s_{_{NN}}} \!=\!$ 200 GeV%
\thanks{Presented at the XXXIII International Symposium on Multiparticle Dynamics}%
}
\author{Patricia Fachini
\address{Brookhaven National Laboratory}
}
\maketitle
\begin{abstract}
The $\rho^0(770)$, $K^{*0}(892)$, $f_{0}(980)$, $\phi(1020)$, and
$\Lambda(1520)$ production in Au+Au and p+p collisions at
$\sqrt{s_{_{NN}}} \!=\!$ 200 GeV are presented. These resonances
are used as a sensitive tool to examine the collision dynamics in
the hadronic medium through their decay and regeneration. The
modification of resonance mass, width, and shape due to phase
space and dynamical effects are also discussed.
\end{abstract}


\section{Introduction}
The measurement of resonances provides an important tool for
studying the dynamics in relativistic heavy-ion collisions by
probing the time evolution of the source from chemical to kinetic
freeze-out and the hadronic interactions at later stages.

Physical effects such as thermal weighting of the states
\cite{1,2,3,4,5,6,7,8} and dynamical interactions with matter
\cite{3,5,7} may modify resonance masses, widths and shapes.
Partial-waves analyses have successfully parameterized $\pi \pi$
scattering \cite{9}. Introducing to the formalism the rescattering
of pions, in which $\pi^{+}\pi^{-} \!\rightarrow\! \rho(770)^0$,
the interference between different scattering channels can distort
the line shape of resonances \cite{10}.


\section{Data Analysis and Results}

The $\rho^0(770)$ \cite{11}, $K^{*0}(892)$ \cite{12}, $f_{0}(980)$
\cite{13}, $\phi(1020)$ \cite{14}, and $\Lambda(1520)$ \cite{15}
production were measured via their hadronic decay channels at
mid-rapidity ($|y| \!\leq\! 0.5$) in Au+Au and p+p collisions at
$\sqrt{s_{NN}}\!=\!$ 200 GeV using the STAR detector at RHIC. The
$\phi$ \cite{16} meson was also measured via its hadronic decay
channel in Au+Au collisions using the PHENIX detector.

\begin{figure}[htb]
\begin{minipage}[t]{60mm}
\begin{center}
\epsfxsize=2.36in \epsfbox{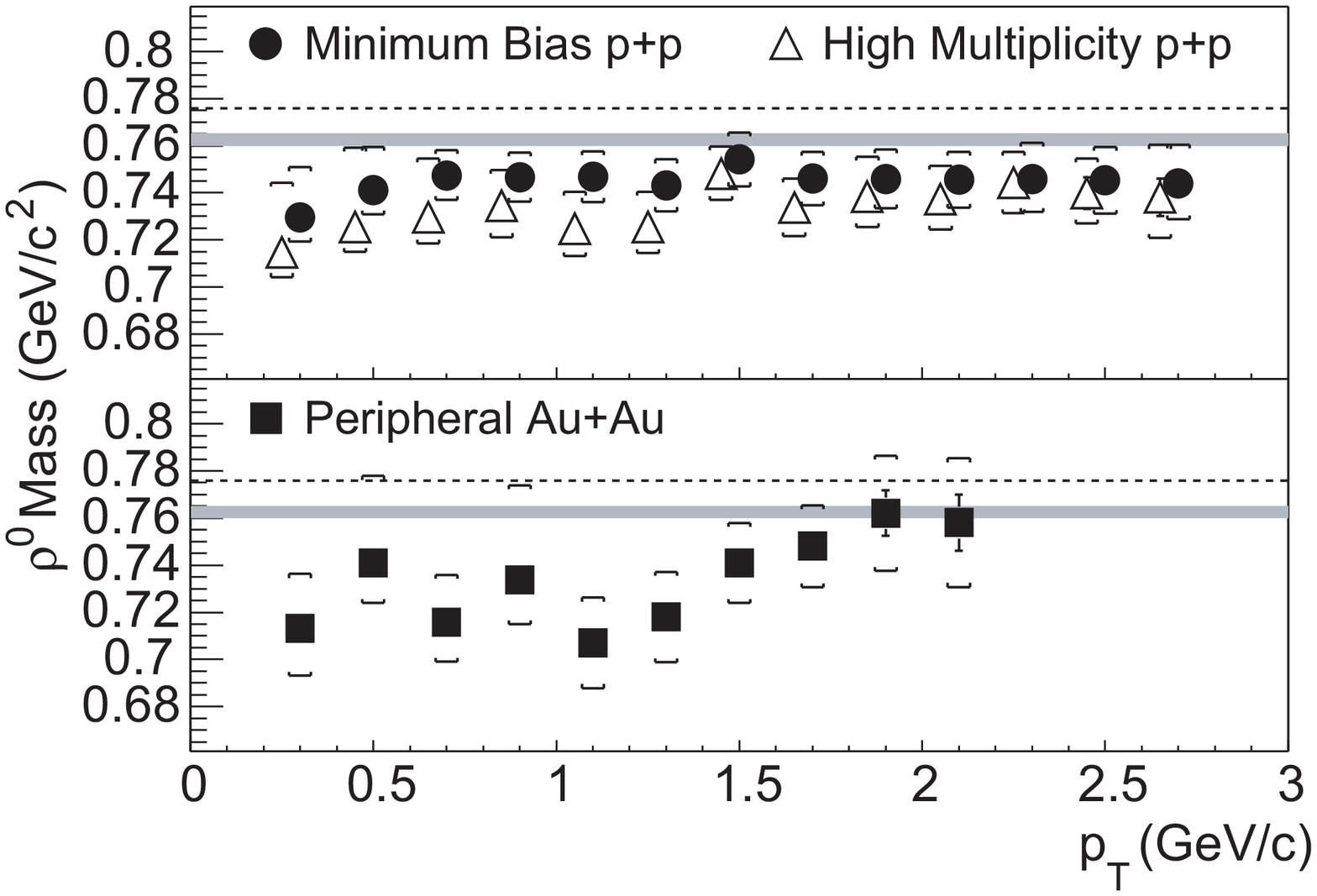}
\end{center}
\end{minipage}
\hspace{\fill}
\begin{minipage}[t]{60mm}
\begin{center}
\epsfxsize=2.31in \epsfbox{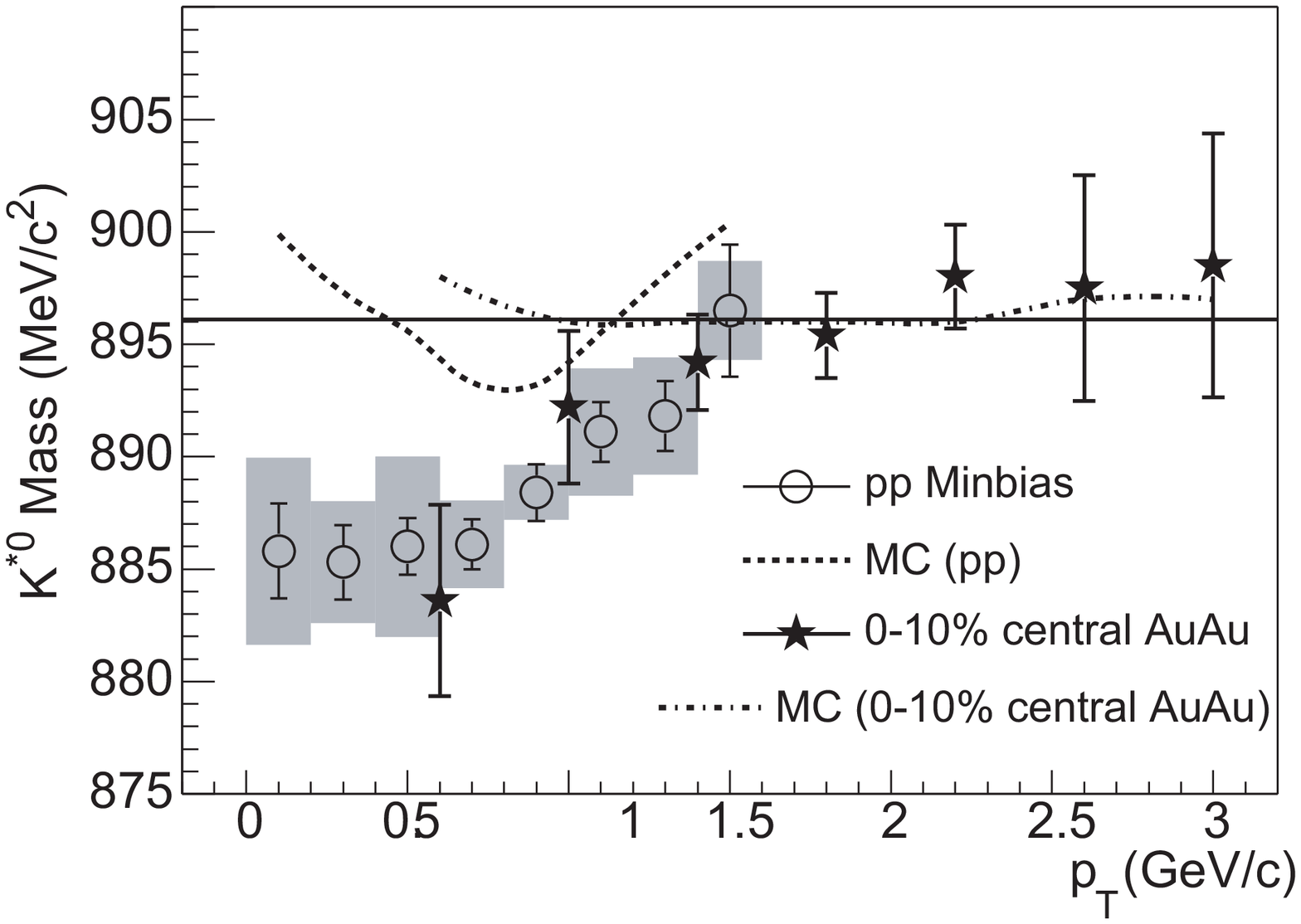}
\end{center}
\end{minipage}
\caption{\label{fig:Mass} Left panel: The $\rho^0$ mass as a
function of $p_T$. The error bars indicate the systematic
uncertainty. The dashed lines represent the average of the
$\rho^0$ mass measured in e$^+$e$^-$ \cite{17}. The shaded areas
indicate the $\rho^0$ mass measured in p+p collisions \cite{18}.
The open triangles have been shifted downward on the abscissa for
clarity. Right panel: $K^{*0}$ mass as a function of $p_T$. The
grey shaded boxes are the systematic uncertainties in minimum bias
p+p. The solid line corresponds to the average of the $K^{*0}$
mass reported in \cite{17}. The dotted and dashed lines are the
results from the Monte Carlo simulations, which accounts for
detector effects and kinematic cuts, for central Au+Au and minimum
bias p+p, respectively.}
\end{figure}

The $\rho^0$ mass is shown as a function of $p_T$ in
Fig.~\ref{fig:Mass} for peripheral Au+Au (40-80$\%$ of the
hadronic cross-section), high multiplicity p+p (top 10$\%$ of the
minimum bias p+p multiplicity distribution for $|\eta| \!<\!$
0.5), and minimum bias p+p interactions. The $\rho^0$ mass was
obtained by fitting the data to a p-wave Breit-Wigner function
times the phase space (BW$\times$PS) described in \cite{11}.
Fig.~\ref{fig:Mass} also shows the $K^{*0}$ mass as a function of
$p_T$ for central Au+Au (top 10$\%$ of the hadronic cross-section)
and minimum bias p+p interactions. The $K^{*0}$ mass was obtained
by fitting the data to the BW$\times$PS function \cite{11} and a
linear function representing the residual background described in
\cite{12}.

The $\rho^0$ and $K^{*0}$ masses increase as a function of $p_T$
and are systematically lower than the value reported in \cite{17}.
The $\rho^0$ mass measured in peripheral Au+Au collisions is lower
than the minimum bias p+p measurement. The $\rho^0$ mass for high
multiplicity p+p interactions is lower than for minimum bias p+p
interactions for all $p_T$ bins, showing that the $\rho^0$ mass is
also multiplicity dependent. Recent calculations are not able to
reproduce the $\rho^0$ mass measured in peripheral Au+Au
collisions without introducing in-medium modification of the
$\rho^0$ meson \cite{3,4,5,6,7,8}.

Previous observations of the $\rho$ meson in e$^+$e$^-$
\cite{19,20,21} and p+p interactions \cite{18} indicate that the
$\rho^0$ line shape is considerably distorted from a p-wave
Breit-Wigner function. A mass shift of $-$30 MeV/$c^{2}$ or larger
was observed in e$^+$e$^-$ collisions at $\sqrt{s}$ $\!=\!$ 90 GeV
\cite{19,20,21}. In the p+p measurement at $\sqrt{s}$ $\!=\!$ 27.5
GeV \cite{18}, a $\rho^0$ mass of 0.7626 $\!\pm\!$ 0.0026
GeV/$c^2$ was obtained from a fit to a relativistic p-wave
Breit-Wigner function times the phase space \cite{18}. This result
is the only p+p measurement used in the hadro-produced $\rho^0$
mass average reported in \cite{17}.

The $\rho^0$ \cite{11} and $f_0$ \cite{13} measurements do not
have sufficient sensitivity to permit a systematic study of the
$\rho^0$ and $f_0$ widths and the $f_0$ mass. The $K^{*0}$
\cite{12} width, the $\phi$ \cite{14,16} mass and width, and the
$\Lambda(1520)$ \cite{15} mass and width are in agreement with the
values reported in \cite{17}.

\begin{figure}[htb]
\begin{minipage}[t]{60mm}
\begin{center}
\epsfxsize=2.35in
\epsfbox{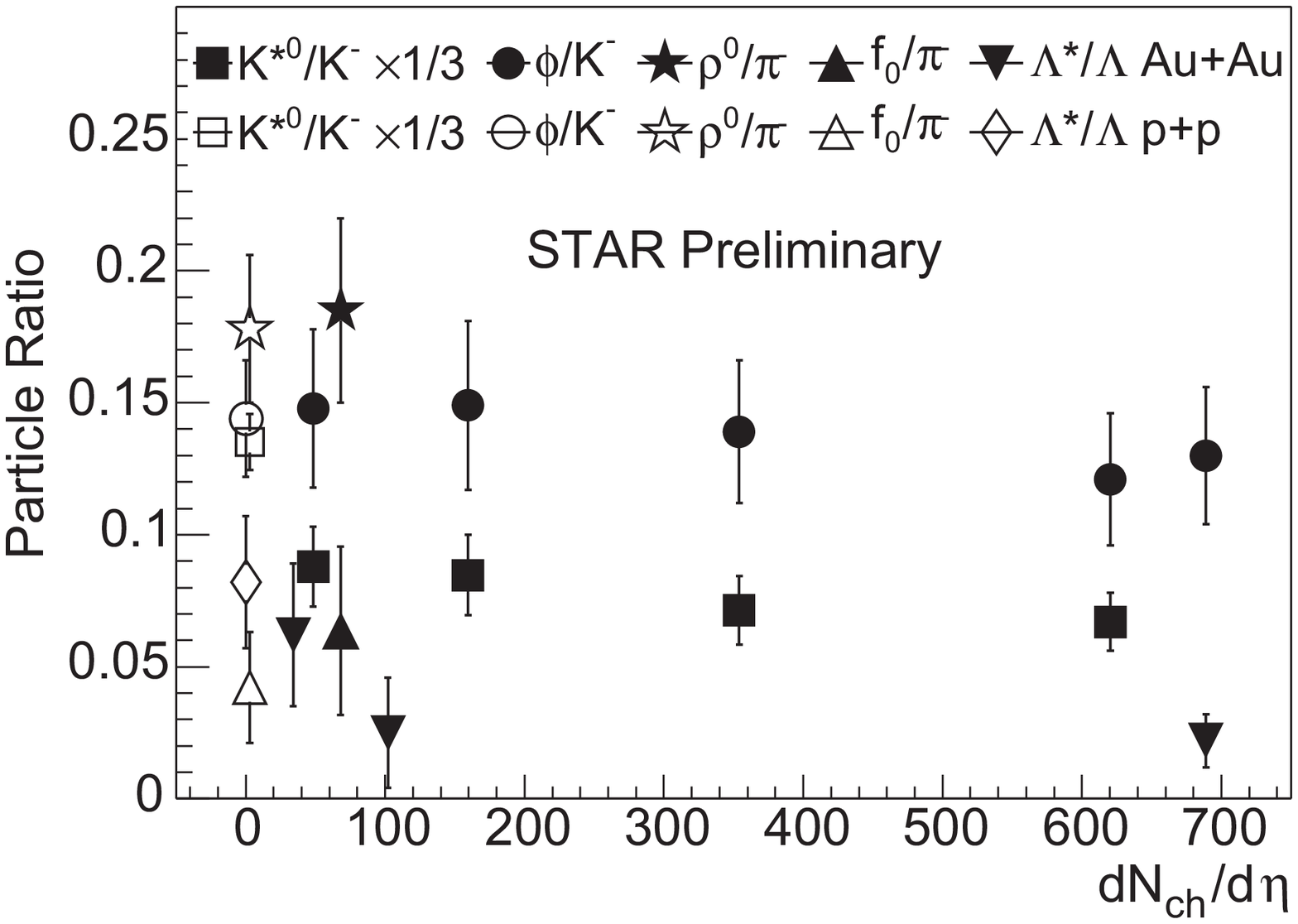}
\end{center}
\end{minipage}
\hspace{\fill}
\begin{minipage}[t]{60mm}
\begin{center}
\epsfxsize=2.35in \epsfbox{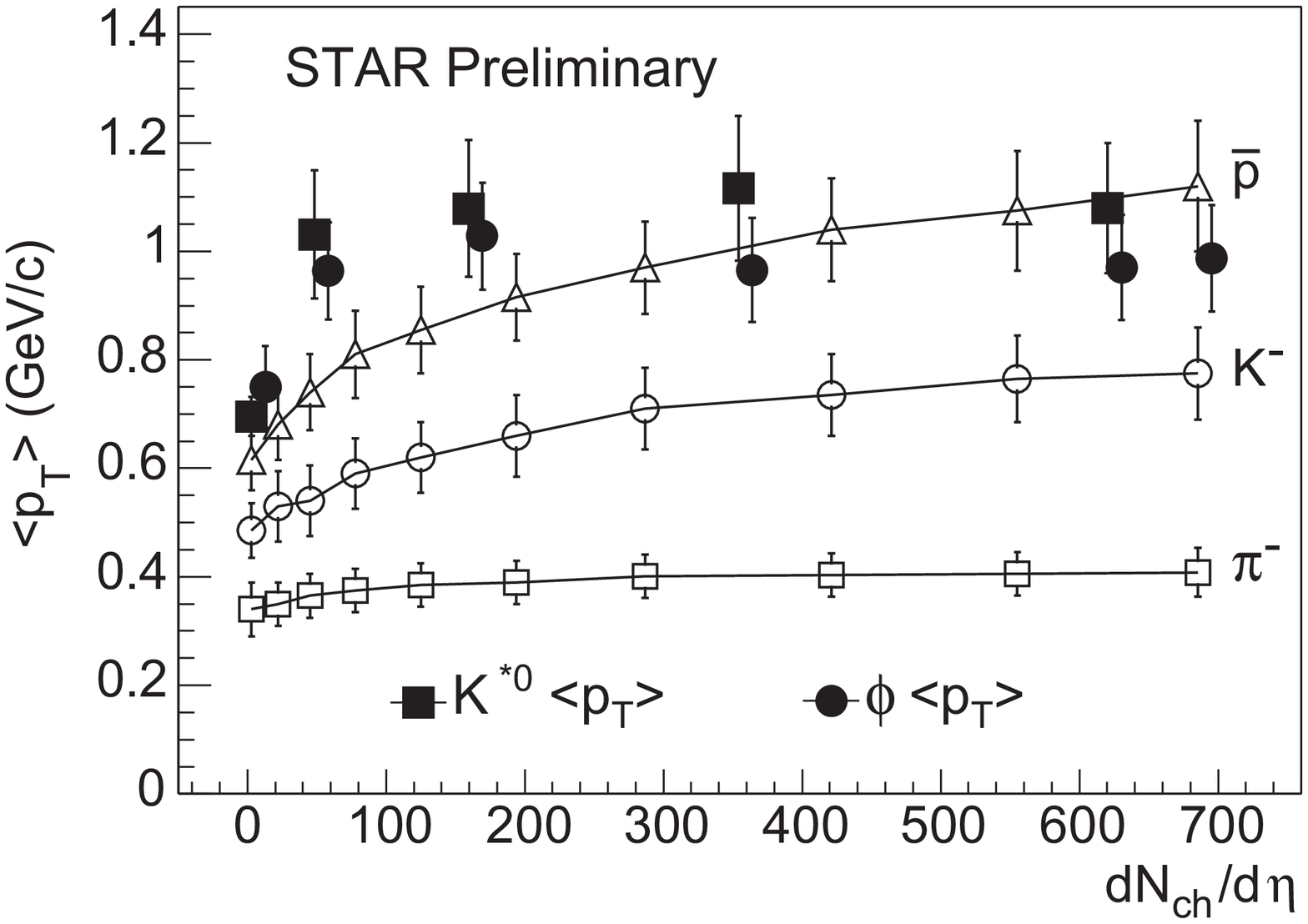}
\end{center}
\end{minipage}
\caption{\label{fig:RatioMeanPt} Left panel: The $\rho^0/\pi^-$,
$K^{*0}/K^-$, $f_{0}/\pi^-$, $\phi/K^-$, and $\Lambda^*/\Lambda$
ratios as a function of $dN_{ch}/d\eta$ for Au+Au and p+p
interactions. Right panel: The $K^{*0}$ and $\phi$ $\langle
p_T\rangle$ as a function of $dN_{ch}/d\eta$ compared to that of
$\pi^-$,$K^-$, and $\bar{p}$.}
\end{figure}

The $\rho^0/\pi^-$, $K^{*0}/K^-$, $f_{0}/\pi^-$, $\phi/K^-$, and
$\Lambda^*/\Lambda$ ratios as a function of charged hadron
multiplicity ($dN_{ch}/d\eta$) for Au+Au and p+p interactions are
depicted in Fig.~\ref{fig:RatioMeanPt}. In the case of the $\phi$,
the effect of the daughter rescattering and regeneration should be
negligible due to the $\phi$ longer lifetime ($\sim$44 fm) and the
small $KK$ cross section \cite{22}, respectively. In the case of
short lived resonances such as the $\rho^{0}$ ($\sim$1.3 fm), the
daughter rescattering and regeneration processes should be
comparable because the $\pi^+\pi^-$ cross-section is dominated by
the $\rho^{0}$ resonance. The $\pi^+\pi^-$ cross-section is larger
than the $K\pi$ cross-section by a factor of $\sim$5 \cite{22}.
Therefore, the daughter rescattering should be the dominant
process in the case of the $K^{*0}$ ($\sim$4 fm). In this picture,
the $\phi/K^-$ and $\rho^0/\pi^-$ ratios should be independent of
$dN_{ch}/d\eta$, while the $K^{*0}/K^-$ ratio should decrease as a
function of $dN_{ch}/d\eta$, which is in agreement with the
measurements presented in Fig.~\ref{fig:RatioMeanPt}.

The $K^{*0}$ and $\phi$ $\langle p_T\rangle$ as a function of
$dN_{ch}/d\eta$ are compared to that of $\pi^-$,$K^-$, and
$\bar{p}$ in Fig.~\ref{fig:RatioMeanPt}. The $K^{*0}$ $\langle
p_T\rangle$ is higher than that of $\bar{p}$ for peripheral
collisions, even though $m_{K^{*0}}\!<\!m_{\bar{p}}$. One
interpretation of this result is that the daughter rescattering is
the dominant process compared to the $K^{*0}$ regeneration, since
only $K^{*0}$ with higher $p_T$ are more likely to decay outside
the fireball. In the case of central Au+Au collisions, the
$K^{*0}$ $\langle p_T\rangle$ is comparable to that of $\bar{p}$
possibly because the hadron production increases and the
regeneration process becomes significant. The $\phi$ $\langle
p_T\rangle$ is independent of $dN_{ch}/d\eta$ in Au+Au collisions.
This is expected if the $\phi$ has a smaller cross-section and
therefore less sensitivity to hadronic rescattering.

\section{Conclusions}
We have presented results on resonance production at mid-rapidity
in Au+Au and p+p collisions at $\sqrt{s_{NN}}$ $\!=\!$ 200 GeV.
The measured $\rho^0$ and $K^{*0}$ masses are $p_T$ dependent and
lower than previous measurements reported in \cite{15}. Dynamical
interactions with the surrounding matter, interference between
various scattering channels, phase space distortions, and
Bose-Einstein correlations are possible explanations for the
apparent modification of resonances properties. The resonances
ratios as a function of $dN_{ch}/d\eta$ may be interpreted in the
context of hadronic cross sections.

\end{document}